\documentclass[prl,aps,
superscriptaddress,twocolumn,preprintnumbers]{revtex4}

\usepackage{graphicx}

\newcommand{\comment}[1]{}

\newcommand{\beq}{\begin{equation}}
\newcommand{\eeq}{\end{equation}}
\newcommand{\bqa}{\begin{eqnarray}}
\newcommand{\eqa}{\end{eqnarray}}
\newcommand{\be}{\begin{equation}}
\newcommand{\ee}{\end{equation}}
\newcommand{\bea}{\begin{eqnarray}}
\newcommand{\eea}{\end{eqnarray}}

\newcommand{\0}{\over }

\newcommand{\6}{\partial }

\newcommand{\tr}{\,{\rm tr}\,}

\newcommand{\sect}[1]{{\em #1.\textemdash}}

\begin{document}

\title{Hard-Loop Dynamics of Non-Abelian Plasma Instabilities }

\preprint{TUW-04-36}
\preprint{HIP-2004-66/TH}
\preprint{BI-TP 2004/37}

\author{Anton Rebhan}
\affiliation{Institut f\"ur Theoretische Physik, Technische Universit\"at Wien,
A-1040 Vienna, Austria}
\author{Paul Romatschke}
\affiliation{Fakult\"at f\"ur Physik, Universit\"at Bielefeld,
D--33501 Bielefeld, Germany}
\author{Michael Strickland}
\affiliation{Institut f\"ur Theoretische Physik, Technische Universit\"at Wien,
A-1040 Vienna, Austria}
\affiliation{
Helsinki Institute of Physics,
P.O. Box 64, FIN-00014 University of Helsinki, Finland}


\begin{abstract}
Non-Abelian plasma instabilities may be responsible for the
fast apparent quark-gluon thermalization in relativistic heavy-ion
collisions if their exponential growth is not hindered by nonlinearities.
We study numerically 
the real-time evolution of instabilities in
an anisotropic 
non-Abelian plasma with an SU(2) gauge group
in the hard-loop approximation.
We find exponential growth of non-Abelian plasma instabilities
both in the linear and in the strongly nonlinear regime, with only
a brief phase of subexponential behavior in between.
\end{abstract}
\pacs{11.15Bt, 04.25.Nx, 11.10Wx, 12.38Mh}

\maketitle

In this Letter we present first results on the 
real-time evolution of non-Abelian plasma 
instabilities due to momentum-space anisotropies in the underlying quark and 
gluon distribution functions
\cite{Stan:9397,Mrowczynski:2000ed,RM:2003,Romatschke:2003ms,Arnold:2003rq,Mrowczynski:2004kv,
Arnold:2004ih} in the nonlinear hard-loop approximation.
Such anisotropies are generated during the natural expansion of the matter 
created during a heavy-ion collision and the resulting
instabilities may be responsible for the fast 
apparent thermalization \cite{Arnold:2004ti},
which seems to be faster than can be accounted for
by perturbative scattering processes 
\cite{Baier:2000sb,Molnar:2001ux,Heinz:2004pj,Shuryak:2004kh}. 
This type of plasma instability is 
the analogue
of  the electromagnetic Weibel instability which causes 
soft gauge (magnetic) fields to become nonperturbatively large.
Eventually this leads to 
large-angle scattering of hard particles \cite{Weibel:1959},
thereby rapidly accelerating the isotropization and subsequent 
thermalization of an Abelian plasma with a temperature anisotropy.
However, in the non-Abelian case it 
is 
conceivable
that the intrinsic nonlinearities 
could cause the instabilities to stop growing 
before they have large effects on hard particles
and therefore reduce their efficacy in isotropizing a non-Abelian plasma.  

The regime where the backreaction of collective fields on the hard
particles is still weak but where the self-interaction of the former
may already be strongly nonlinear is governed by a
``hard-loop'' effective action which
has been derived in Ref.~\cite{Mrowczynski:2004kv}
for arbitrary momentum-space anisotropies \footnote{In
Ref.~\cite{Mrowczynski:2004kv} the distribution
function is assumed to be constant in space and time.
This is certainly not the most general situation
imaginable in real heavy-ion collisions. However,
the growth rate of the plasma instabilities
is expected to be such that the time dependence
due to expansion of a realistic plasma can be neglected \cite{Arnold:2004ti}.}.
We discretize this effective
action in a local auxiliary-field formulation, 
keeping its full 
dynamical nonlinearity and nonlocality.
This is then applied to initial
conditions that allow for an effectively 1+1-dimensional lattice simulation,
extending a previous numerical study \cite{Arnold:2004ih} that
used a static and linear approximation to the hard-loop effective action.

\sect{Discretized Hard-Loop Dynamics}%
At weak gauge coupling $g$, there is a separation of
scales in hard momenta $|\mathbf p|=p^0$ of (ultrarelativistic)
plasma constituents, and soft momenta $\sim g|\mathbf p|$
pertaining to collective dynamics.
The effective field theory for the soft modes that is
generated by integrating out the hard plasma modes
at one-loop order and in the approximation that
the amplitudes of the soft gauge fields obey $A_\mu \ll |\mathbf p|/g$
is that of gauge-covariant collisionless Boltzmann-Vlasov equations
\cite{HTLreviews}.
In equilibrium, the corresponding (nonlocal) effective action is
the so-called hard-thermal-loop effective action 
\cite{HTLeffact
} which has a simple generalization
to plasmas with anisotropic momentum distributions \cite{Mrowczynski:2004kv}.
Its contribution to the effective field equations of soft modes
is an induced current of the form \cite{Blaizot:1993be,Mrowczynski:2000ed}
\begin{equation}\label{Jind0}
j^\mu[A] = -g^2
\int {d^3p\over(2\pi)^3} 
{1\over2|\mathbf p|} \,p^\mu\, {\partial f(\mathbf p) \over \partial p^\beta}
W^\beta(x;\mathbf v),
\end{equation}
where $f$ is a 
weighted sum of the quark and gluon distribution 
functions \cite{Mrowczynski:2004kv}.
The quantities $W_\beta(x;\mathbf v)$ satisfy
\begin{equation}\label{Weq}
[v\cdot D(A)]W_\beta(x;\mathbf v)
= F_{\beta\gamma}(A) v^\gamma \, 
\end{equation}
\comment{mpt}
with $v^\mu\equiv p^\mu/|\mathbf p|=(1,\mathbf v)$, 
$D_\mu=\partial_\mu-ig[A_\mu,\cdot]$, metric
convention $(+---)$,
and this has to be solved
self-consistently with
$D_\mu(A) F^{\mu\nu}=j^\nu$.
At the expense of having introduced a continuous set of auxiliary fields 
$W_\beta(x;\mathbf v)$ the effective field equations are local, but nonlinear in
the case of a non-Abelian gauge theory.

We shall be interested in particular in the case where
there is just one direction of momentum-space anisotropy (e.g.,
the collision axis in heavy-ion experiments), so we assume
cylindrical symmetry and parametrize the first derivatives of
the hard particle distribution
function by 
\bea
{\partial f(\mathbf p) \over \partial p^\beta}
&=&
{\partial f(|\mathbf p|,p^z) \over \partial |\mathbf p|}
{p^b \delta_{b\beta}\0|\mathbf p|}
+{\6 f(|\mathbf p|,p^z) \over \6 p^z}\delta_{z\beta}\nonumber\\
&\equiv&
-\tilde f_1(|\mathbf p|,p^z)
{p^b \delta_{b\beta}\0|\mathbf p|}
-\tilde f_2(|\mathbf p|,p^z)\delta_{z\beta}.
\eea
\comment{mpt}

Because $p^\beta W_\beta=0$, we have
\be\label{Jind2}
j^\mu(x)={1\02}g^2
\int {d^3p\over(2\pi)^3} 
v^\mu\, [\tilde f_1 W^0(x;v)+\tilde f_2W^z(x;v)].
\ee
\comment{mpt}

In the isotropic case one has $\tilde f_2=0$, and only $W^0$
appears, whose equation of motion (\ref{Weq}) is driven by (chromo)electric
fields; in the anisotropic case, however, $W^z$ enters, whose
equation of motion involves the $z$ component of the 
Lorentz force.

The induced current (\ref{Jind0}) is covariantly conserved
as one can verify by partial
integration with respect to $p^\beta$
\cite{Mrowczynski:2004kv}, but 
no partial integration is needed for
parity invariant distribution
functions $f$. This is only a mild restriction
on the choice of $f$'s, but
a great advantage for the
following discretization. For $f(\mathbf p)=f(-\mathbf p)$,
the two terms in the current (\ref{Jind2}) are covariantly 
conserved individually, because Eq.~(\ref{Weq})
implies that
$D\cdot j \propto
\int\! {d^3p}\, 
(\tilde f_1 F^{0\gamma}v_\gamma+\tilde f_2 F^{z\gamma}v_\gamma)$, which vanishes by symmetry.

Our aim will be to study the hard-loop dynamics
in lattice discretization where we approximate the continuous
set of fields $W_\beta(x;\mathbf v)$ by a finite number of
fields.
Because Eq.~(\ref{Weq}) does not mix different $v$'s, a closed set of
equations is obtained when the integral in Eq.~(\ref{Jind2})
is discretized with respect to directions $\mathbf v$
\bea\label{Jindd}
j^\mu(x)&=&g^2
\int {p^2 dp\over(2\pi)^2} {1\0\mathcal N} \sum_{\bf v} 
v^\mu\, [\tilde f_1 W^0_{\mathbf v}(x)+
\tilde f_2 W_{\mathbf v}^z(x)]\nonumber\\
&&\equiv {1\0\mathcal N} \sum_{\bf v} v^\mu[a_{\mathbf v} W^0_{\mathbf v}(x)+
b_{\mathbf v} W^z_{\mathbf v}(x)],
\eea
where the $\mathcal N$ unit vectors $\mathbf v$ define a partition
of the unit sphere in patches of equal area, and where
$a_{\mathbf v}$, $b_{\mathbf v}$
are then fixed coefficients for a given distribution function $f$.
Covariant conservation of $j$ is ensured
by symmetric choices of the set of
$\mathbf v$'s which satisfy $\sum_{\mathbf v} a_{\mathbf v} \mathbf v = 0$,
$\sum_{\mathbf v} b_{\mathbf v}  = 0$, and
$\sum_{\mathbf v} b_{\mathbf v} \mathbf v_\perp  = 0$,
with $\mathbf v_\perp = \mathbf v - v^z \mathbf e_z$.


Given such a discretization, which is a discretization
of the phase space of the hard
particles with respect to the directions of their
momenta, only $\mathcal N$ auxiliary fields
$\mathcal W_{\mathbf v}=a_{\mathbf v} W^0_{\mathbf v}+
b_{\mathbf v} W^z_{\mathbf v}$ participate in the dynamical
evolution. The full hard-loop dynamics is then approximated
by the following set of matrix-valued equations,
\bea\label{DHLW}
&&[v\cdot D(A)]\mathcal W_{\mathbf v}=(a_{\mathbf v} F^{0\mu}+b_{\mathbf v}
F^{z\mu})v_\mu\\
\label{DHLF}
&&D_\mu(A) F^{\mu\nu}=j^\nu={1\0\mathcal N}\sum\limits_{\mathbf v} v^\nu 
\mathcal W_{\mathbf v},
\eea
which can be systematically improved by increasing $\mathcal N$.

A different possibility for discretizing hard-loop dynamics that
has been employed previously in isotropic plasmas would
be a decomposition of the auxiliary fields $W^\mu(x;v)$
into spherical harmonics
\cite{Bodeker:1999gx,Rajantie:1999mp}. Our proposal is slightly simpler, but also more
flexible in that it allows one to e.g.\ improve approximations in highly
anisotropic cases with cylindrical symmetry by selectively
increasing the resolution in $z$ direction more than in $\varphi$
direction.

The dynamical system described by Eqs.~(\ref{DHLW}) and
(\ref{DHLF}) has constant total energy of the form
\be\label{Etot}
\mathcal E=
\int d^3x \tr (\mathbf E^2 + \mathbf B^2)|_t + 
\int_{t_0}^t\! dt' \int d^3x\, 2 \tr \mathbf j_{t'}\cdot \mathbf E_{t'}.
\ee
The part containing the induced current
involves
\bea
\tr \mathbf j\cdot \mathbf E
&=&{g^2\04}\6_\mu \int {d^3p\over(2\pi)^3} \tilde f_1 v^\mu W_0^2\nonumber\\
&&+{g^2\02}\int {d^3p\over(2\pi)^3} \tilde f_2 W^z (v\cdot D)W_0,
\eea
which shows that in the isotropic case ($\tilde f_2=0$) there is
a local, positive definite energy contribution from the plasma
\cite{Blaizot:1994am}.
However, in the general anisotropic case, positivity is lost, corresponding
to the possibility of plasma instabilities, where
energy may be extracted from hard particles and deposited into
the soft collective fields without bound
as long as the hard-loop approximation $A_\mu \ll |\mathbf p|/g$
remains valid.


{\em 1+1-dimensional lattice simulation.---}%
When only $z$-dependent initial conditions are imposed,
the entire dynamics proceeds 1+1-dimensionally and we can
take all collective fields as 1+1-dimensional, though
the underlying hard degrees of freedom are of course still 3+1 dimensional,
with a discrete set of directions $\mathbf v$ for their (conserved)
momenta.
This dimensionally reduced situation already allows us to study
the evolution of non-Abelian
Weibel instabilities, which in the linear (Abelian) regime
are formed by (superpositions of)
transverse standing waves with exponentially growing amplitudes. 

We have solved the equations (\ref{DHLW}) and (\ref{DHLF}) 
in this dimensionally reduced situation
by lattice discretization, closely following Ref.~\cite{Arnold:2004ih}
who considered a toy model corresponding to an induced current
which is simply
\be\label{jAL}
j_\alpha^{\rm AL}=\mu^2 A_\alpha,\quad \alpha=x,y.
\ee
As was shown in Ref.~\cite{Arnold:2004ih}, this reproduces
the static limit of the anisotropic hard-loop effective
action for fields that vary only in the anisotropy direction, 
but it neglects its general frequency dependence and dynamical
nonlinearity. This is now provided
through Eqs.~(\ref{DHLW})
and (\ref{DHLF}).

Like Ref.~\cite{Arnold:2004ih} we work in temporal
axial gauge $A^0=0$ and take initial conditions corresponding to
small random chromoelectric fields $\mathbf E=-\6_t \mathbf A$
with polarization transverse to the $z$-axis,
and all other fields vanishing, which
in our case includes the auxiliary fields
$\mathcal W_{\mathbf v}$. 
This initial condition satisfies the Gauss law constraint,
$
\mathbf D \cdot \mathbf E = {\mathcal N}^{-1}\sum_{\mathbf v} 
\mathcal W_{\mathbf v},
$
whose continued fulfilment is monitored in the simulation, but not
enforced. As a further nontrivial check of our numerics,
we tracked conservation of
the total energy (\ref{Etot}).

In order to be able to compare with the analytical results of
Ref.~\cite{Romatschke:2003ms
}, we
consider an anisotropic distribution function 
$f=N(\xi) f_{iso}(\mathbf p^2+\xi p_z^2)$. This determines
the coefficients in Eq.~(\ref{Jindd}) according to
\be\label{avbvxi}
a_{\mathbf v}={m^2 (1+\xi v_z^2)^{-2} },\qquad
b_{\mathbf v}=\xi v_z a_{\mathbf v},
\ee
where $m^2=N(\xi)m_D^2(0)$ with
$m_D(0)$ the Debye mass of the isotropic case $\xi=0$,
and $N(\xi)$ a normalization factor
which equals $N(\xi)=\sqrt{1+\xi}$ if one requires
that the number density of hard particles
remains the same for different values $\xi$.

\begin{figure}
\includegraphics[width=2.45in]{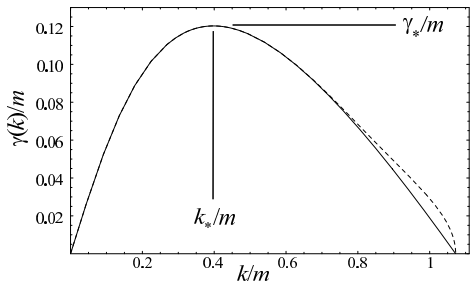}
\caption{The growth rate $\gamma(k)$ of unstable modes for the discretized
hard loops with
$\mathcal N=100$ (dashed line)
in comparison with the continuum result (solid line)
for $\xi=10$.
\label{figgamma}}
\end{figure}

For $\xi>0$ transverse modes with wave vector $|\mathbf k|<\mu$,
\be
\mu^2={1\04}\left(1+{\xi-1\0\sqrt{\xi}}\arctan{\sqrt\xi}\right)m^2,
\ee
are unstable and grow with a $k$-dependent growth rate
$\gamma(k)$, shown in
Fig.~\ref{figgamma} for $\xi=10$, which has a maximum $\gamma_*$
at nonvanishing
$|\mathbf k|=k_*$ and a zero at $\mathbf k=0$.
By contrast, the growth rate implied by the toy model (\ref{jAL})
is given by $\gamma=\sqrt{\mu^2-\mathbf k^2}$, whose maximum is
at $\mathbf k=0$ and equals $\mu$.

In the linearized case one can also determine the dispersion laws analytically
in the case of finite $\mathcal N$. 
Fig.~\ref{figgamma} compares the growth rates $\gamma(k)$
for full and discretized hard loops with
$\mathcal N=\mathcal N_z \times\mathcal N_\varphi=20\times5=100$,
which shows that the latter give very accurate approximations
for $\mathcal N\gtrsim 100$.
As another example, the asymptotic mass of the stable propagating
modes with $|\mathbf k| \gg \mu$ is given for full hard loops
by $m^2_\infty={m^2/(2\sqrt{\xi})}\arctan{\sqrt\xi}$,
which for $\xi=10$, $\mathcal N=100$ is reproduced with an error
of only 0.017\%.

\begin{figure}
\includegraphics[width=2.6in]{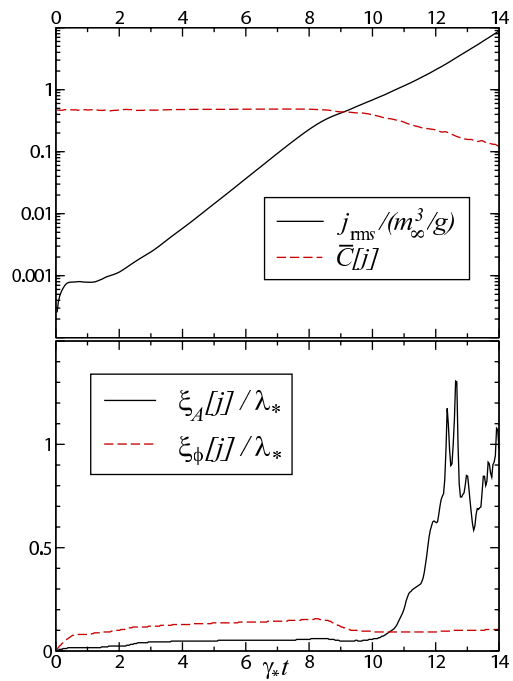}
\caption{The average current $j_{rms}$ and the average
relative size $\bar C$ of commutators $[j_x,j_y]$ as a function of time
(upper panel);
the Abelianization correlation length $\xi_A[j]$ and the ordinary
correlation length $\xi_\phi[j]$, normalized
to the scale of maximal growth $\lambda_*$ (lower panel).
\label{figN}}
\end{figure}

\begin{figure}
\includegraphics[width=2.84in]{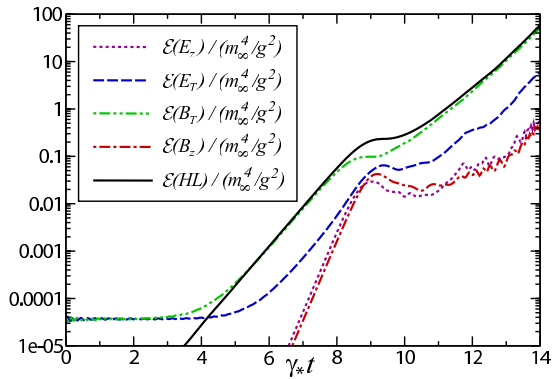}
\caption{Average energy densities $\mathcal E$
in transverse/longitudinal chromomagnetic/electric
fields and the total energy density contributed by
hard particles, $\mathcal E({\rm HL})$.
\label{figNen}}
\end{figure}

The results of our lattice simulations of the
nonlinear evolution for a gauge group SU(2) are finally shown in
Figs.\ \ref{figN} and \ref{figNen}, where we have used the parameters
$\xi=10$, $\mathcal N_z\times\mathcal N_\varphi=20\times 5$
for discretizing the unit sphere
with uniform spacing in $z$ and $\varphi$.
The one-dimensional spatial
lattice has 10,000 sites, periodic boundary conditions,
and lattice spacing corresponding 
to $am_\infty=0.028$ so that the physical size
is $L\approx 280\,m_\infty^{-1}$. 
Using a leapfrog algorithm with
time steps $\epsilon/a=1/100$,
we track the evolution of a single field configuration 
\footnote{We have convinced ourselves of its generic nature
by evolving, on various lattices, a few dozen configurations
with randomly varying seed fields and different $\mathcal N$.}
with
random initial seed chromoelectric field of
root-mean-square amplitude $0.012\,m_\infty^2/g$.
The Gauss law constraint turns out to remain preserved
within machine accuracy; the total
energy (\ref{Etot}) is conserved within less than 1\%, with the percentage
level being reached
only at the largest times when
the energy in the soft fields has grown by more than
a factor of $10^6$
(full details will be given elsewhere). 

The upper panel of Fig.~\ref{figN} shows the evolution of 
\be
j_{rms}=
\left[ \int_0^L {dz\0L} 2 \tr (\mathbf j^2) \right]^{1/2},
\ee
which (after some initial
wobble) grows exponentially with a growth rate that is most of the
time only slightly below $\gamma_*$, 
except for a transitory reduction at the beginning
of the nonlinear regime, when $j_{rms}\sim m_\infty^3/g$. 
Also shown is the dimensionless
observable $\bar C$, defined by \footnote{%
This definition coincides with the definition in
Ref.~\cite{Arnold:2004ih} in the simplifying case (\ref{jAL}), but
is gauge invariant also when the restriction to 1+1 dimensional
configurations is removed.}
\be
\bar C = \int_0^L {dz\0L} { \left\{ \tr \left( (i[j_x,j_y])^2 \right) 
\right\}^{1/2}\0 \tr (j_x^2+j_y^2) }
\ee
and giving a measure of local ``non-Abelianness''. In the toy model
of Ref.~\cite{Arnold:2004ih} it was found that $\bar C$ suddenly begins to
decay exponentially when fields get strong,
with a decay rate similar to the growth rate of $j_{rms}$.
In the hard-loop case, we observe a similar phenomenon, but the
decay rate of $\bar C$ is much smaller (and also less constant
once the decay begins).

Ref.~\cite{Arnold:2004ih} also observed a concurrent global
Abelianization, by comparing the correlation among parallel
transported spatially separated commutators to the correlation
of parallel transported fields. A correlation length $\xi_A$
defined through the former was found to rapidly grow to lattice size
when $\bar C$ begins to decay, whereas no such growth occurred in
the general field correlation length $\xi_\phi$. In
the lower panel of Fig.~\ref{figN}
we show the analogous quantities computed using currents instead
of field values [22] 
and we found that correlated Abelianization takes place
over extended domains, which remain bounded, however. In fact, $\xi_A$
turns out to be comparable with 
the scale of maximal growth
$\lambda_*=2\pi/k_* \approx 7m_\infty^{-1}$.
By contrast, 
in the
model of Ref.~\cite{Arnold:2004ih} $k_*$ vanishes,
which is presumably responsible for the different global behavior.

Fig.~\ref{figNen} shows how the exponentially growing energy transferred
from hard to soft scales is distributed among chromomagnetic
and chromoelectric collective fields. The dominant contribution
is in transverse magnetic fields, and it grows roughly with
the maximum rate $\gamma_*$ both in the linear as well as in
the highly nonlinear regime, with a transitory slowdown in between. 
Transverse electric fields behave similarly,
and are suppressed by a factor of the order of $(\gamma_*/k_*)^2$
\footnote{In the 
model of Ref.~\cite{Arnold:2004ih} the situation
is again different: Because $k_*$ is zero there, the dominant
energy component is from transverse electric fields, whereas the
relative importance of magnetic fields drops with time.}.
The appearance of
longitudinal contributions, which are absent in the initial conditions
we have chosen, is a purely non-Abelian effect. While 
completely negligible at first,
they have a growth rate which is double the one in the transverse sector,
and they begin to catch up with the latter just when local
Abelianization sets in. At that stage there appears to be some
complicated rearrangement taking place, which delays
the exponential evolution by a time
$\Delta t\sim \gamma_*^{-1}$, but subsequently the growth
rate gets restored roughly to its previous value. 
Eventually there will be a point where the hard-loop
approximation breaks down, namely when the energy transferred
from hard to soft modes becomes comparable to that initially
present in the former.

From the above numerical results on the hard-loop dynamics of non-Abelian
plasma instabilities we conclude that the latter do not saturate
until they begin to have large effects on hard particle trajectories,
even though our simulations indicate complicated dynamics and only
limited Abelianization of the unstable modes.
Therefore it appears indeed
possible that non-Abelian plasma instabilities are responsible for
accelerated thermalization in a weakly coupled quark-gluon plasma.

\acknowledgments

We 
are indebted to
Peter Arnold for
correspondence on 
Ref.~\cite{Arnold:2004ih},
and we also thank D.~B\"odeker, M.~Laine, T.~Lappi, G.~Moore,
and K.~Rummukainen
for useful discussions. M.S. was supported by the Austrian Science 
Fund FWF, project no.\ M790, and by the Academy of Finland, contract no. 77744.



\end{document}